\documentclass[prl,reprint]{revtex4-1}

\usepackage{amsmath}
\usepackage{bm}
\usepackage{amssymb}
\usepackage{graphicx}
\usepackage{hyperref}
\newcommand{\normord}[1]{:\mathrel{#1}:}
\begin{document}
\title{Universality of photon counting below a bifurcation threshold}

\author{Lisa Arndt}
\email{lisa.arndt@rwth-aachen.de}
\affiliation{JARA Institute for Quantum Information, RWTH Aachen University, 52056 Aachen, Germany}

\author{Fabian Hassler}
\affiliation{JARA Institute for Quantum Information, RWTH Aachen University, 52056 Aachen, Germany}

\date{September 2020}
\begin{abstract} At a bifurcation point, a small change of a parameter causes
a qualitative change in the system. Quantum fluctuations wash out this abrupt
transition and enable the emission of quantized energy, which we term photons,
below the classical bifurcation threshold. Close to the bifurcation point, the
resulting photon counting statistics is determined by the instability. We
propose a generic method to derive a characteristic function of photon
counting close to a bifurcation threshold that only depends on the dynamics
and the type of bifurcation, based on the universality of the
Martin-Siggia-Rose action. We provide explicit expressions for the cusp
catastrophe without conservation laws.  Moreover, we propose an experimental
setup using driven Josephson junctions that exhibits both a fold and a
pitchfork bifurcation behavior close to a cusp catastrophe.

\end{abstract}
\maketitle

Universality is a central theme in modern statistical mechanics. A prime
example is the universality of the critical exponents that describe the
divergence of observables close to a second-order phase transition
\cite{chaikin:95,stanley:99,pelissetto:02}.  Catastrophe theory offers
insights into the universality as it categorizes how small changes in external
parameters can lead to qualitative changes in the behavior of the system
\cite{arnold:86}.  As such, it also provides a suitable framework to study
phase transitions in driven-dissipative systems
\cite{haken:75,strogatz:00,carusotto:13,ritsch:13}.  In these system, finite
frequency excitations are studied---which we call photons in the following. At
the bifurcation threshold, a small change of the system parameters leads to a
condensation of these photons \cite{levin:05}. Typical examples are the lasing
\cite{svelto:10} and the Dicke transition \cite{kirton:19} in optical systems.

In the vicinity of a bifurcation point, a characteristic long time-scale
emerges in the dissipative dynamics of the system. As a result, the
qualitative properties of phase transitions can be described only by a small
number of relevant degrees of freedom exhibiting the slow dynamics. Moreover,
a (quasi-)classical treatment of the dynamics is appropriate since the number
of photons becomes large in the vicinity of the bifurcation threshold. Note
that the dynamics of slow classical degrees of freedoms has been grouped into
universality classes by Halperin and Hohenberg \cite{hohenberg:77}.

Different types of bifurcations and dynamics in driven-dissipative systems
lead to a variety of critical exponents and correlation behaviors that have
been the focus of many studies in recent years
\cite{torre:13,sieberer:13,brennecke:13,raftery:14,dagvadorj:15,nagy:16,sibalic:16,marino:16,biondi:17,comaron:17,hwang:18,young:20}.
So far, mirroring the discussion of equilibrium physics, these exponents are
derived for low-order cumulants of physical observables. Note that in
equilibrium physics, the central limit theorem leads to a Gaussian statistics
of all relevant observables and moreover the fluctuations are connected to the
amount of dissipation. On the other hand, in driven-dissipative systems the
fluctuation-dissipation theorem does not hold and even more importantly
non-Gaussian statistics should be expected in general which leads to the
question of determining the critical exponents of higher order cumulants.

Motivated by this insight, we employ a path integral formalism
\cite{pilgram:03,elgart:04,jordan:04,jordan:04b,Padurariu:12,chantasri:15} to
investigate the photon counting statistics below the bifurcation threshold. We
find that the critical exponents of all the cumulants only depend on the type
of bifurcation given a Halperin-Hohenberg dynamics. In particular, this result
explains that the full counting statistics of the degenerate
\cite{Vyas:89,Padurariu:12} and non-degenerate parametric oscillator
\cite{Vyas:92,arndt:19} are equivalent close to the threshold, as was noted in
Ref.~\cite{arndt:19}. Based on the universality of the Martin-Siggia-Rose
action, we propose a generic method to derive a universal characteristic
function of photon counting close to a bifurcation threshold. We exemplify our
formalism for a cusp catastrophe with a system dynamics without conservation
laws. We note that the universality of counting statistics in non-equilibrium systems
has been discussed before in different contexts, \emph{e.g.}, for the
statistics of topological defects \cite{campo:18,gomez:20}, particle transport
\cite{duarte:13,baek:19}, and non-equilibrium fluctuation theorem
\cite{Crooks:99,esposito:09}.

The article is organized as follows. Our starting point is the classical
description of the bifurcation dynamics by the corresponding
Martin-Siggia-Rose action. From there, we derive a universal expression for
the characteristic function of photon counting below the bifurcation threshold
by including a normal-ordered counting term. To demonstrate the formalism, we
analyze the counting statistics and rare event statistics for a fold and a
pitchfork bifurcation within a cusp catastrophe framework. Finally, we propose
a microwave experiment that can demonstrate the critical exponents in the higher
order cumulants of the cusp catastrophe.

Each model in the Halperin-Hohenberg classification can be mapped to a
corresponding classical Martin-Siggia-Rose action $S_\text{MSR}$. In the
following, we want to consider photon radiation with a linewidth $\Gamma$ and
frequency $\Omega\gg\Gamma$ that emanates from a system close to a
driven-dissipative phase transition. For simplicity, we focus on a single
harmonic oscillator whose `slow' dynamics in the rotating frame is purely
dissipative. In the Halperin-Hohenberg classification, this dynamics
corresponds to universality class A also known as the Glauber model. Our
results apply mutatis mutandis to dissipative field theories which are relevant
to lattices of coupled cavities, see \emph{e.g.} Refs.~\cite{schmidt:09,carusotto:09,marino:16}.

For our system, the Martin-Siggia-Rose action of the dimensionless `slow'
variable $x$ is given by $S_\mathrm{MSR}(x,\tilde{x},\beta)=\int
dt\,\tilde{x}[\dot{x}+\Gamma x- f(x)+\frac{i}{2} \beta\Gamma \tilde{x}]$ with
the response field $\tilde{x}$ which satisfies the commutation relation
$[x,\tilde x]=i$. This action corresponds to a Langevin equation of the form $
\dot{x}= -\Gamma x+ f(x)+\xi(t)$. The Gaussian fluctuations $\xi(t)$ have the
correlations $\langle \xi(t)\rangle=0$ and $\langle \xi(0) \xi(t)\rangle=
\beta\Gamma \delta (t)$, with a classical, temperature-dependent noise
parameter $\beta\propto k_B T/\hbar\Omega$. The term $-\Gamma x$ corresponds to the
dissipative part of the force in the rotating frame. The remaining force
$f(x)=-V'(x)$ is non-dissipative and includes the external driving force.

In the quantum description of the system, a parameter $\alpha$ is introduced
that relates the quadrature $x^2$ to the photon number $n = x^2/\alpha$. In
this sense, $\alpha$ plays the role of $\hbar$. We are interested in the
quasi-classical regime with $\alpha\ll1$ where the quantum fluctuations remain
small \cite{schmid:82,kleinert:95}.   We relate the variable $\tilde x$ to the
quantum scale by introducing the conjugate variable $p=\alpha\tilde{x}$ with the canonical
commutation relation $[x,p]=i \alpha$.  Additionally, as we are interested in the quantum regime
with $k_BT\ll\hbar\Omega$, where fluctuations are dominated by quantum effects, we
have to replace $\beta$ by $\alpha$ \footnote{Note that
at finite temperatures, we have $\beta \mapsto \alpha (1 +2 n_B)$ with $n_B =
(e^{\hbar\Omega/k_B T} -1)^{-1}$ and we obtain the classical limit with $k_B T
\gg \hbar
\Omega$.}.
The characteristic function of the
photon counting statistics can be obtained by adding a source term to the
Martin-Siggia-Rose action such that the generating function of the photon
counting statistics is given by $\mathcal{Z}(\chi) =
\int\!\mathcal{D}[x]\mathcal{D}[p]\,\exp[iS(\chi)]$ with 
\begin{multline}\label{eq:sq}
  S(\chi) =
  S_\mathrm{MSR}(x,\alpha^{-1}p,\alpha )
  + 
   \frac{\chi \Gamma}{\alpha}
   \!\int_0^\tau \!dt\,\left(x + \tfrac{i}2 \alpha p\right)^{2}  \,,
\end{multline}
where $\tau$ is the detection time. The generating function
$\mathcal{Z}(\chi)$ represents the characteristic function of the number of
photons $N$ detected within  the  time $\tau$; from this, all the cumulants
can be obtained via $\langle\!\langle N^k\rangle\!\rangle=d^k
\ln(\mathcal{Z})/d(i\chi)^k|_{\chi=0}$.

The special form of the last term in Eq.~\eqref{eq:sq} is one of the main results
of our work.  It originates from the normal-ordering of the photon number
operator $\hat{n}=\hat{x}^2/\alpha$ and is a pure quantum effect. It can be
understood as follows: using the conventional creation and annihilation
operators $\hat{b}^\dag$ and $\hat{b}$, the normal-ordered operator is of the
form $\normord{\hat{x}}\,=(\alpha/2)^{1/2}(\hat{b}^\dagger_-+\hat{b}_+)$, with
the notation $\hat{O}_+(\rho)=\hat{O}\rho$ and $\hat{O}_-(\rho)=\rho\hat{O}$
where $\rho$ is the density operator. An equivalent way of writing the
normal-ordered operator is given by
$\normord{\hat{x}}\,=\hat{x}_s+\tfrac{i}{2} \alpha\hat{p}_a$, with
$\hat{x}_s=\tfrac12(\hat{x}_++\hat{x}_-)$, $\hat{p}_a=\hat{p}_+-\hat{p}_-$, and
$\hat{p}=i(\alpha/2)^{1/2}(\hat{b}^\dagger-\hat{b})$. Since the operators
$\hat{x}_s$ and $\hat{p}_a$ are canonically conjugate with
$[\hat{x}_s,\hat{p}_a]=i \alpha$, they correspond to the variables $x$ and $p$
in the quasi-classical action.  Note that the normal ordering leads to the
additional terms $i \alpha xp - \tfrac14 \alpha^2 p^2$ when expanding
$(x+\tfrac{i}2 \alpha p)^2$ that are coupled to the counting field but vanish
in the classical limit $\alpha\to 0$ (at fixed $x, \tilde x$).

In the first part of this paper, we have shown how the characteristic function
close to bifurcation can be obtain purely from the knowledge of the classical
Martin-Siggia-Rose action. In the following, we exemplify this method by
calculating the counting statistics for the explicit choice of the potential
$V(x)/\Gamma=x^4-ax^2/2+bx$, with the dimensionless parameters $a$ and $b$
\footnote{Note that the dissipative part of the force can be included via the
substitution $a\mapsto a-1$.}. This potential corresponds to a cusp
catastrophe which enables us to study the two most fundamental bifurcations:
the fold and the pitchfork bifurcation. In the following, we elaborate on and
demonstrate the introduced method by calculating explicitly the critical
exponents and rare-event statistics for both types of bifurcations.

\begin{figure}[tb]
	\centering
	\includegraphics{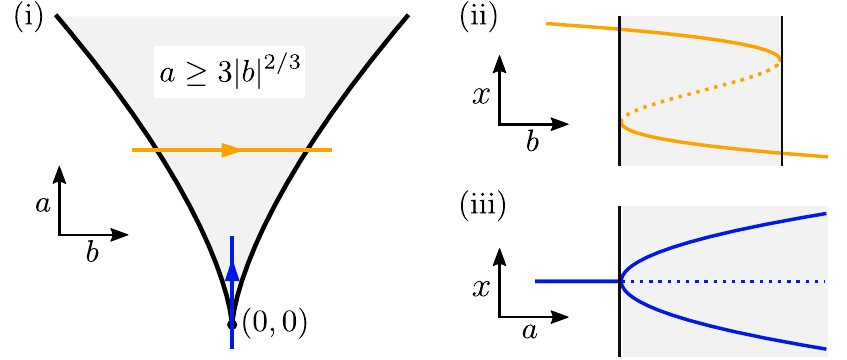}
	\caption{%
		(i) Cusp shape in the canonical variables $(a,b)$. The fold
              bifurcation lines at $b=\pm(a/3)^{3/2}$ separate the region
            outside the cusp, where a single stable solution exists, from the
          region inside the cusp (gray) where two stable solutions exist. (ii) Evolution of the fixed points as the systems passes through the cusp along the horizontal line indicated in (i). At the first fold bifurcation, a second stable solution (solid line) and an unstable point (dotted line) become available, while the stability of the upper point remains unchanged. The classical switch between both solutions takes place at the second fold bifurcation, where the upper fixed point and the unstable fixed point merge forcing the system onto the lower, stable fixed point. (iii) Evolution of the fixed points as the systems passes to $a>0$ through the cusp point $(0,0)$ along the vertical line indicated in (i). At the pitchfork bifurcation, the single stable fixed point splits into two stable (solid lines) and one unstable (dotted line) solution.}\label{fig:cusp}
\end{figure}

In the parameter space $(a,b)$, the system can be divided into two regions: the first region, outside the cusp, where a single stable solution exists and the second region, inside the cusp, where two stable solutions exist. The cusp shape is shown in Fig.~\ref{fig:cusp}(i). It is formed by the lines $b=\pm(a/3)^{3/2}$ for $a\geq 0$.

In the symmetrical case $b=0$, the system displays a pitchfork bifurcation as
it passes to $a>0$ through the cusp point $(0,0)$. Here, one stable, classical
solution splits into two stable solutions and one unstable solution as
displayed in Fig.~\ref{fig:cusp}(iii). Away from the cusp point, the system
displays a fold bifurcation, where in addition to the single stable solution
that already exists outside the cusp an alternate second solution that is
paired with an unstable solution becomes available. Disregarding fluctuations,
the classical state of the system does not change upon this transition due to
hysteresis. Instead, the classical switch between both solutions takes place at the second fold bifurcation as indicated in Fig.~\ref{fig:cusp}(ii). Here, the initial classical solution vanishes forcing the system onto the second fixed point.

First, we want to calculate the counting statistics in the vicinity of the
pitchfork bifurcation at $(0,0)$, with $\delta=-a>0$ the distance from the
bifurcation point. For $b=0$, this corresponds to the counting statistic of the
parametric oscillator which has already been analyzed in
Refs.~\cite{Padurariu:12,arndt:19}. In the quasi-classical limit $\alpha\ll1$,
vacuum phase fluctuations remain small. Upon approaching the threshold, the
fluctuations increase. However, for sufficiently small $\alpha$, the crossover
region where the fluctuations become of the order of $1$ remains narrow and can
be estimated as $\delta\simeq \alpha$. It is therefore valid to expand the
Martin-Siggia-Rose action to quadratic order around the stationary solution below the instability threshold. For $|b|\ll \delta$, the stationary solution is given by $x=p=0$. 
 
We focus on the limit of long measurement times $\Gamma\tau\gg 1$ and calculate the cumulant-generating function $\lambda(z=i\chi)=\ln[\mathcal{Z}(\chi)]$. In the vicinity of the threshold, we obtain the result
\begin{align}\label{eq:lambdacusp}
\lambda(z)=\frac{\Gamma\tau}{2}\left[\delta-\sqrt{\delta^2-2z}+\frac{2b^2z}{\alpha(\delta^2-2z)}\right]
\end{align}
which corresponds to the cumulants
\begin{align}\label{eq:momentscusp}
\frac{\langle\!\langle N^k\rangle\!\rangle}{\Gamma
\tau}=\frac{(2k-3)!!}{2\,\delta^{2k-1}}+\frac{2^{k-1}\,k!\,b^2}{\alpha\,\delta^{2k}}.
\end{align}
The counting statistics has two distinct contributions: The first term is due to
the pure pitchfork
statistics at $b=0$ which has been previously reported in
Refs.~\cite{Padurariu:12,arndt:19}. The second term is due to the contribution of
the fold at finite $b$. Note that the cumulants diverge as $\delta^{-\gamma_k}$
with $\gamma_k$ the critical exponent of the $k$-th cumulant. The pitchfork
yields $\gamma_k=2k-1$ which, for $k=1$, reproduces the critical exponent of
the number of photons in the cavity in a Dicke model discussed in
Ref.~\cite{kirton:19}. The fold contribution demonstrates an even stronger
divergence with $\gamma_k=2k$. At finite but small $b$, we thus predict a
crossover behavior with the critical exponent changing from $2k-1$ to $2k$ when
approaching the threshold \cite{chaikin:95}. The
Fano factor $F=\langle\!\langle N^2\rangle\!\rangle/\langle N\rangle$ is a
measure of the number of correlated photons. For the pure pitchfork statistic
this factor is given by $F=\delta^{-2}\propto n^2$ with $n=\langle
N\rangle/2\Gamma\tau$ the average number of photons in the system. Thus, the
number of correlated photons exceeds by far the number of photons present in
the system at any given time \cite{Padurariu:12}. The photons are thus
correlated over the long, divergent time scale $\tau^*=
F/\bar{I}=2/\Gamma\delta$, with $\bar{I}=\langle N\rangle/\tau$ the average
photon current. This is a central characteristic of the bifurcation behavior.
The divergent time scale is also responsible for the large deviations from the
average photon current. In particular, the probability to measure a photon current $I$ during the measurement time $\tau$ is given by 
\begin{align}\label{eq:extremecusp}
P(I)\propto\exp\left[-\frac{\tau}{2\tau^*}\left(\frac{\bar{I}}{I}+\frac{I}{\bar{I}}-2\right)\right],
\end{align}
up to exponential accuracy \cite{Padurariu:12,arndt:19}. Note that the
probability distribution is strongly asymmetric with the probability to
measure a current smaller than average currents strongly suppressed when
compared to the Gaussian approximation.

Next, we want to compare the results of the pitchfork to the fold bifurcation
which is the instability  away from the cusp point. In this regime, two
saddle points contribute to the counting statistics which correspond to the
two stable solutions above the transition. Note that the second saddle point
is at a finite value of the quantum variable $p$ since there is
only a single stable classical solution below the transition.
Analogously to the calculation at the cusp point, we want to avoid the crossover region close to the bifurcation point where phase
fluctuation increase without bounds. For the fold bifurcation, this narrow region can be
estimated as $\delta\simeq \alpha^2b^{-4/3}$, where  we introduced the measure
$\delta=3 b^{2/3}(3 b^{2/3}-a)>0$ for the distance to the
bifurcation \footnote{Since the bifurcation is symmetric in $b$, we simplify
the notation by focusing on the case $b>0$}.  Away from this region, both
saddle points are well separated and the total cumulant-generating function is
given by $\lambda(z)=\ln
[p_1e^{\lambda_1(z)}+p_2e^{\lambda_2(z)}]$, with
$\lambda_{1,2}(z)$ the cumulant-generating function of the first/second saddle
point. The probabilities $p_{j}$, with $p_1 + p_2=1$, denote the fraction of
time the $j$-saddle point contributes to the total counting
statistics \cite{pilgram:03}. We find that the second saddle
point  is exponentially suppressed as compared to the first saddle point with
$p_2\propto 
\exp(-\Gamma\tau\delta^{1/2})$. This is
due to the fact that the second saddle point is at a finite value of the quantum
variable such that it is
only probed by rare quantum fluctuations. 

\begin{figure}[tb]
	\centering
	\includegraphics{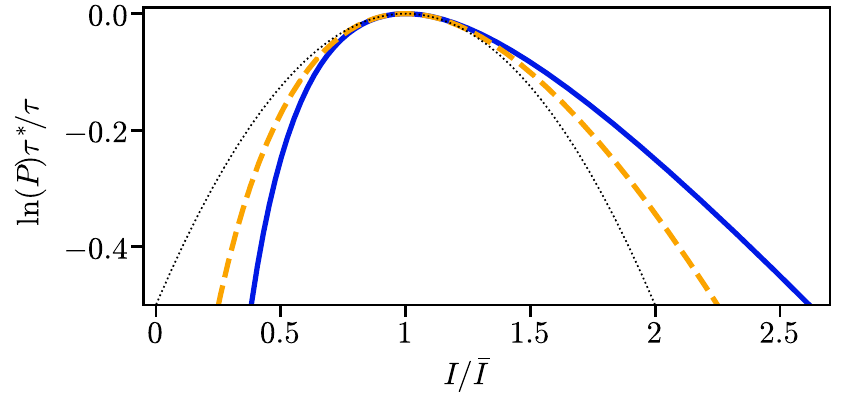}
	\caption{%
		Probability of large deviations of the photon current $I$ from
              the average current $\bar{I}$ in the vicinity of the pitchfork
            bifurcation (blue, solid line) and the fold bifurcation (orange,
          dashed line). The scaling is chosen such that both probabilities have
        identical Gaussian expansions around $I=\bar{I}$ as indicated by the
      thin, dotted line. Note that both probabilities are strongly asymmetric leaning
      towards large current deviations. }\label{fig:extreme}
\end{figure}

Evaluating $\lambda_{1,2}$, we find that only the second saddle point leads to
a divergent counting statistics. Because of this, we require intermediate
times, $1\ll\Gamma \tau\ll\delta^{-1/2}$, to observe the critical exponents
such that $p_2 \approx p_1 \approx \frac12$ and the relevant part of the
counting statistics is given by $\lambda(z)=\frac12\lambda_2(z)$. Note that at
longer measurement times, the probability $p_2$ to be at the second saddle
point decreases exponentially reducing the prefactor of the divergences.
At intermediate time scales, we obtain the result
\begin{align}\label{eq:lambdafold}
\lambda(z)=\frac{\Gamma\tau b^{2/3}\delta z}{8\alpha(\delta-z)}
\end{align}
for the leading order behavior in $\delta$ and $b$ close to the bifurcation
threshold with the cumulants
\begin{align}\label{eq:momentsfold}
\frac{\langle\!\langle N^k\rangle\!\rangle}{\Gamma \tau}=\frac{k!\,b^{2/3}}{8
\alpha\,\delta^{k-1}}.
\end{align}
In this case, the critical exponents $\gamma_k=k-1$  are different from the pitchfork
bifurcation. In particular, the photon
current at the fold does not show any divergence and only the Fano factor $F=2
\delta^{-1}$ shows a divergence. The diverging time-scale in this case is given
by $\Gamma\tau^*=16 \alpha b^{-2/3} \delta^{-1}$ and only diverges as
$\delta^{-1}$ when compared to $\delta^{-2}$ before. Note that the condition
$\delta\gtrsim \alpha^2b^{-4/3}$ from above makes sure that we are in the
intermediate regime with $\Gamma \tau^* \lesssim \delta^{-1/2}$. The
rare-event statistics is given by
\begin{align}\label{eq:extremefold}
P(I)\propto\exp\left[-\frac{2\tau}{ \bar{I}\tau^*}\Bigl(
I^{1/2}-\bar{I}\,^{1/2}\Bigr)^2\right].
\end{align}
In Fig.~\ref{fig:extreme}, we compare the probability of large deviation of
both types of bifurcations. Compared to the Gaussian approximation, both types
of bifurcation show an increased probability for larger photon currents. At
fixed $\bar I \tau^*$, the pitchfork leads to larger fluctuations
with $|\ln P|_\text{pitchfork} \approx \tfrac14 |\ln P|_\text{fold}$.
For deviations below the average current, we observe that for the fold
bifurcation the probability to observe a current $I \to 0$ is finite
$P_\text{fold}(I \to 0) \propto \exp(-2 \tau/\tau^*) $ whereas
the corresponding probability vanishes in the pitchfork case.

Before concluding, we want to discuss possible experimental realizations in
which the predicted exponents can be observed. The behavior of a Dicke
transition \cite{kirton:19} at finite number of spins can be mapped to a
pitchfork bifurcation. Such a system is realized with cold-atomic gases in
a cavity \cite{ritsch:13}. Measuring the statistics of the photons that are leaking out of the
cavity will enable a comparison to the predicted critical exponents. The paradigmatic
example of  the fold bifurcation is the laser transitions (in rotating wave)
\cite{haken:75,strogatz:00,fiedler:08} whose physical realizations are
ubiquitous. We predict the counting statistics of the photons
below the lasing transition to be universal and to follow 
Eq.~\eqref{eq:momentsfold}. In the following, we highlight an implementation using
superconducting circuits where the full cusp catastrophe, in particular the
crossover from pitchfork to the fold transition, can be observed. The setup
extends the circuit of  Ref.~\cite{Padurariu:12} with an additional ac-current source to
account for the asymmetry in the cusp potential. The total setup is composed of
a Josephson junction with Josephson energy $E_J$, biased by a dc-voltage
source, that is in series with a microwave resonator and an ac-current source.
The resonator is characterized by a resonance frequency $\Omega$, the decay
rate $\Gamma$, and an impedance $Z_0$ at low frequency. 
To observe the cusp catastrophe, we set the frequency of the current source to
the resonance frequency $\Omega$ with $I(t)=I_0\sin (\Omega t)$ and tune the
Josephson frequency to twice the resonance frequency by setting the dc-bias
voltage to $V=\hbar\Omega/e$. Additionally, we require the phase between both drives
to be fixed. We assume that the impedance far from resonance $Z_0$ is small at
the quantum scale such that the vacuum fluctuation strength
$\alpha=8e^2 Z_0/\hbar\ll1$ \footnote{We also want to neglect the broadening of
  the Josephson emission line by low-frequency phase noise. It can be
  neglected, if it is much smaller than the linewidth of the resonator
$\hbar\Gamma\gg \alpha k_B T $ \cite{likharev,Arndt:18}.}.
The calculation of the corresponding action follows similar steps as the
derivations in Refs.~\cite{Padurariu:12,arndt:19}, \emph{i.e.}, starting
with a Keldysh path integral formalism and performing a rotating wave approximation. The
final step is a projection of the resulting action onto the `slow' direction of
the dynamics in the vicinity of the cusp bifurcation line. It yields the
Martin-Siggia-Rose action of the Glauber model for the force
$f(x)/\Gamma=8\epsilon J_2(x)/x-j$ with the Bessel functions $J_{m}(x)$ and the
parameters $j= \alpha I_0/4e\Gamma$ and $\epsilon= \alpha E_J/4\hbar\Gamma$. 
To discuss the behavior in the vicinity of the cusp catastrophe, it is
sufficient to expand the Bessel function to fourth order in $x$. Then, the
force is given by $f(x)/\Gamma=-\epsilon x^3/12+\epsilon x- j$, which leads to
a cusp bifurcation line at $j=\pm \frac43\epsilon^{-1/2}(\epsilon-1)^{3/2}$ for
$\epsilon\gtrsim1$. For the counting statistic in the vicinity of the pitchfork bifurcation, the mapping to the previous parameters is straightforward with $\delta=1-\epsilon$ and $b=j$. For the fold bifurcation, we obtain the mapping $\delta=(\frac34j)^{2/3}[(\frac34j)^{2/3}-\epsilon+1]$ and $b=48j$ to leading order in $j$.

In conclusion, we have outlined a method to derive the universal characteristic
function of photon counting close to a bifurcation threshold. While our results focused on the Glauber model without any spatial dependence, our approach to derive the characteristic function from the classical Martin-Siggia-Rose action can be easily mapped to other models or problems with spatial dependence for which other universality classes can be studied. The most important step is the exchange of the classical counting term by its normal-ordered quantum equivalent.
We have demonstrated the proposed method by calculating the photon counting
statistics below the cusp threshold for the fold as well as the pitchfork
bifurcation. Superficially, both bifurcations lead to a divergent counting statistics upon
approaching the bifurcation threshold. However, the critical exponents
$\gamma_k$ as well as the  probabilities of rare events differ. Possible
ways to test the universal statistics include the lasing \cite{fiedler:08} and
the Dicke transition \cite{kirton:19}. Additionally, we have proposed a
microwave setup based on the degenerate parametric oscillator
\cite{Padurariu:12} that exhibits a cusp catastrophe and could thus be used to
observe both sets of critical exponents in a single device.


\begin{thebibliography}{10}
\makeatletter

\bibitem{chaikin:95}
P. M. Chaikin and T. C. Lubensky,
 {\em Principles of Condensed Matter Physics\/}
 (Cambridge University Press, 1995).

\bibitem{stanley:99}
H. E. Stanley,
 Scaling, universality, and renormalization: Three pillars of modern critical
  phenomena,
 \href{http://dx.doi.org/10.1103/RevModPhys.71.S358}{%
 Rev. Mod. Phys. {\bf 71}, S358 (1999)}.

\bibitem{pelissetto:02}
A. Pelissetto and E. Vicari,
 Critical phenomena and renormalization-group theory,
 \href{http://dx.doi.org/https://doi.org/10.1016/S0370-1573(02)00219-3}{%
 Physics Reports {\bf 368} (6), 549  (2002)}.

\bibitem{arnold:86}
R. Thomas and V. Arnol'd,
 {\em Catastrophe Theory\/}
 (Springer Berlin Heidelberg, 1986).

\bibitem{haken:75}
H. Haken,
 Cooperative phenomena in systems far from thermal equilibrium and in
  nonphysical systems,
 \href{http://dx.doi.org/10.1103/RevModPhys.47.67}{%
 Rev. Mod. Phys. {\bf 47}, 67 (1975)}.

\bibitem{strogatz:00}
S. H. Strogatz,
 {\em Nonlinear Dynamics and Chaos: With Applications to Physics, Biology,
  Chemistry and Engineering\/}
 (Westview Press, 2000).

\bibitem{carusotto:13}
I. Carusotto and C. Ciuti,
 Quantum fluids of light,
 \href{http://dx.doi.org/10.1103/RevModPhys.85.299}{%
 Rev. Mod. Phys. {\bf 85}, 299 (2013)}.

\bibitem{ritsch:13}
H. Ritsch, P. Domokos, F. Brennecke, and T. Esslinger,
 Cold atoms in cavity-generated dynamical optical potentials,
 \href{http://dx.doi.org/10.1103/RevModPhys.85.553}{%
 Rev. Mod. Phys. {\bf 85}, 553 (2013)}.

\bibitem{levin:05}
M. A. Levin and X.-G. Wen,
 String-net condensation: A physical mechanism for topological phases,
 \href{http://dx.doi.org/10.1103/PhysRevB.71.045110}{%
 Phys. Rev. B {\bf 71}, 045110 (2005)}.

\bibitem{svelto:10}
O. Svelto,
 {\em Principles of Lasers\/}
 (Springer US, 2010).

\bibitem{kirton:19}
P. Kirton, M. M. Roses, J. Keeling, and E. G. Dalla~Torre,
 Introduction to the dicke model: From equilibrium to nonequilibrium, and vice
  versa,
 \href{http://dx.doi.org/10.1002/qute.201800043}{%
 Advanced Quantum Technologies {\bf 2} (1-2), 1800043 (2019)}.

\bibitem{hohenberg:77}
P. C. Hohenberg and B. I. Halperin,
 Theory of dynamic critical phenomena,
 \href{http://dx.doi.org/10.1103/RevModPhys.49.435}{%
 Rev. Mod. Phys. {\bf 49}, 435 (1977)}.

\bibitem{torre:13}
E. G. D. Torre, S. Diehl, M. D. Lukin, S. Sachdev, and P. Strack,
 Keldysh approach for nonequilibrium phase transitions in quantum optics:
  Beyond the dicke model in optical cavities,
 \href{http://dx.doi.org/10.1103/PhysRevA.87.023831}{%
 Phys. Rev. A {\bf 87}, 023831 (2013)}.

\bibitem{sieberer:13}
L. M. Sieberer, S. D. Huber, E. Altman, and S. Diehl,
 Dynamical critical phenomena in driven-dissipative systems,
 \href{http://dx.doi.org/10.1103/PhysRevLett.110.195301}{%
 Phys. Rev. Lett. {\bf 110}, 195301 (2013)}.

\bibitem{brennecke:13}
F. Brennecke, R. Mottl, K. Baumann, R. Landig, T. Donner, and T. Esslinger,
 Real-time observation of fluctuations at the driven-dissipative dicke phase
  transition,
 \href{http://dx.doi.org/10.1073/pnas.1306993110}{%
 Proceedings of the National Academy of Sciences {\bf 110} (29), 11763 (2013)}.

\bibitem{raftery:14}
J. Raftery, D. Sadri, S. Schmidt, H. E. T\"ureci, and A. A. Houck,
 Observation of a dissipation-induced classical to quantum transition,
 \href{http://dx.doi.org/10.1103/PhysRevX.4.031043}{%
 Phys. Rev. X {\bf 4}, 031043 (2014)}.

\bibitem{dagvadorj:15}
G. Dagvadorj, J. M. Fellows, S. Matyja\ifmmode~\acute{s}\else
  \'{s}\fi{}kiewicz, F. M. Marchetti, I. Carusotto, and M. H.
  Szyma\ifmmode~\acute{n}\else \'{n}\fi{}ska,
 Nonequilibrium phase transition in a two-dimensional driven open quantum
  system,
 \href{http://dx.doi.org/10.1103/PhysRevX.5.041028}{%
 Phys. Rev. X {\bf 5}, 041028 (2015)}.

\bibitem{nagy:16}
D. Nagy and P. Domokos,
 Critical exponent of quantum phase transitions driven by colored noise,
 \href{http://dx.doi.org/10.1103/PhysRevA.94.063862}{%
 Phys. Rev. A {\bf 94}, 063862 (2016)}.

\bibitem{sibalic:16}
N. \ifmmode \check{S}\else \v{S}\fi{}ibali\ifmmode~\acute{c}\else \'{c}\fi{},
  C. G. Wade, C. S. Adams, K. J. Weatherill, and T. Pohl,
 Driven-dissipative many-body systems with mixed power-law interactions:
  Bistabilities and temperature-driven nonequilibrium phase transitions,
 \href{http://dx.doi.org/10.1103/PhysRevA.94.011401}{%
 Phys. Rev. A {\bf 94}, 011401 (2016)}.

\bibitem{marino:16}
J. Marino and S. Diehl,
 Quantum dynamical field theory for nonequilibrium phase transitions in driven
  open systems,
 \href{http://dx.doi.org/10.1103/PhysRevB.94.085150}{%
 Phys. Rev. B {\bf 94}, 085150 (2016)}.

\bibitem{biondi:17}
M. Biondi, G. Blatter, H. E. T\"ureci, and S. Schmidt,
 Nonequilibrium gas-liquid transition in the driven-dissipative photonic
  lattice,
 \href{http://dx.doi.org/10.1103/PhysRevA.96.043809}{%
 Phys. Rev. A {\bf 96}, 043809 (2017)}.

\bibitem{comaron:17}
P. Comaron, G. Dagvadorj, A. Zamora, I. Carusotto, N. P. Proukakis, and M. H.
  Szyma\ifmmode~\acute{n}\else \'{n}\fi{}ska,
 Dynamical critical exponents in driven-dissipative quantum systems,
 \href{http://dx.doi.org/10.1103/PhysRevLett.121.095302}{%
 Phys. Rev. Lett. {\bf 121}, 095302 (2018)}.

\bibitem{hwang:18}
M.-J. Hwang, P. Rabl, and M. B. Plenio,
 Dissipative phase transition in the open quantum rabi model,
 \href{http://dx.doi.org/10.1103/PhysRevA.97.013825}{%
 Phys. Rev. A {\bf 97}, 013825 (2018)}.

\bibitem{young:20}
J. T. Young, A. V. Gorshkov, M. Foss-Feig, and M. F. Maghrebi,
 Nonequilibrium fixed points of coupled ising models,
 \href{http://dx.doi.org/10.1103/PhysRevX.10.011039}{%
 Phys. Rev. X {\bf 10}, 011039 (2020)}.

\bibitem{pilgram:03}
S. Pilgram, A. N. Jordan, E. V. Sukhorukov, and M. B\"uttiker,
 Stochastic path integral formulation of full counting statistics,
 \href{http://dx.doi.org/10.1103/PhysRevLett.90.206801}{%
 Phys. Rev. Lett. {\bf 90}, 206801 (2003)}.

\bibitem{elgart:04}
V. Elgart and A. Kamenev,
 Rare event statistics in reaction-diffusion systems,
 \href{http://dx.doi.org/10.1103/PhysRevE.70.041106}{%
 Phys. Rev. E {\bf 70}, 041106 (2004)}.

\bibitem{jordan:04}
A. N. Jordan, E. V. Sukhorukov, and S. Pilgram,
 Fluctuation statistics in networks: A stochastic path integral approach,
 \href{http://dx.doi.org/10.1063/1.1803927}{%
 Journal of Mathematical Physics {\bf 45} (11), 4386 (2004)}.

\bibitem{jordan:04b}
A. N. Jordan and E. V. Sukhorukov,
 Transport statistics of bistable systems,
 \href{http://dx.doi.org/10.1103/PhysRevLett.93.260604}{%
 Phys. Rev. Lett. {\bf 93}, 260604 (2004)}.

\bibitem{Padurariu:12}
C. Padurariu, F. Hassler, and Y. V. Nazarov,
 Statistics of radiation at josephson parametric resonance,
 \href{http://dx.doi.org/10.1103/PhysRevB.86.054514}{%
 Phys. Rev. B {\bf 86}, 054514 (2012)}.

\bibitem{chantasri:15}
A. Chantasri and A. N. Jordan,
 Stochastic path-integral formalism for continuous quantum measurement,
 \href{http://dx.doi.org/10.1103/PhysRevA.92.032125}{%
 Phys. Rev. A {\bf 92}, 032125 (2015)}.

\bibitem{Vyas:89}
R. Vyas and S. Singh,
 Photon-counting statistics of the degenerate optical parametric oscillator,
 \href{http://dx.doi.org/10.1103/PhysRevA.40.5147}{%
 Phys. Rev. A {\bf 40}, 5147 (1989)}.

\bibitem{Vyas:92}
R. Vyas,
 Photon-counting statistics of the subthreshold nondegenerate parametric
  oscillator,
 \href{http://dx.doi.org/10.1103/PhysRevA.46.395}{%
 Phys. Rev. A {\bf 46}, 395 (1992)}.

\bibitem{arndt:19}
L. Arndt and F. Hassler,
 Statistics of radiation due to nondegenerate josephson parametric
  down-conversion,
 \href{http://dx.doi.org/10.1103/PhysRevB.100.014505}{%
 Phys. Rev. B {\bf 100}, 014505 (2019)}.

\bibitem{campo:18}
A. del Campo,
 Universal statistics of topological defects formed in a quantum phase
  transition,
 \href{http://dx.doi.org/10.1103/PhysRevLett.121.200601}{%
 Phys. Rev. Lett. {\bf 121}, 200601 (2018)}.

\bibitem{gomez:20}
F. J. G\'omez-Ruiz, J. J. Mayo, and A. del Campo,
 Full counting statistics of topological defects after crossing a phase
  transition,
 \href{http://dx.doi.org/10.1103/PhysRevLett.124.240602}{%
 Phys. Rev. Lett. {\bf 124}, 240602 (2020)}.

\bibitem{duarte:13}
G. C. Duarte-Filho, F. A. G. Almeida, S. Rodr\'{\i}guez-P\'erez, and A. M. S.
  Mac\^edo,
 Charge counting statistics and weak localization in a quantum chain,
 \href{http://dx.doi.org/10.1103/PhysRevB.87.075404}{%
 Phys. Rev. B {\bf 87}, 075404 (2013)}.

\bibitem{baek:19}
Y. Baek, Y. Kafri, and V. Lecomte,
 Finite-size and finite-time effects in large deviation functions near
  dynamical symmetry breaking transitions,
 \href{http://dx.doi.org/10.1088/1742-5468/ab43d5}{%
 Journal of Statistical Mechanics: Theory and Experiment {\bf 2019} (10),
  103202 (2019)}.

\bibitem{Crooks:99}
G. E. Crooks,
 Entropy production fluctuation theorem and the nonequilibrium work relation
  for free energy differences,
 \href{http://dx.doi.org/10.1103/PhysRevE.60.2721}{%
 Phys. Rev. E {\bf 60}, 2721 (1999)}.

\bibitem{esposito:09}
M. Esposito, U. Harbola, and S. Mukamel,
 Nonequilibrium fluctuations, fluctuation theorems, and counting statistics in
  quantum systems,
 \href{http://dx.doi.org/10.1103/RevModPhys.81.1665}{%
 Rev. Mod. Phys. {\bf 81}, 1665 (2009)}.

\bibitem{schmidt:09}
S. Schmidt and G. Blatter,
 Strong coupling theory for the jaynes-cummings-hubbard model,
 \href{http://dx.doi.org/10.1103/PhysRevLett.103.086403}{%
 Phys. Rev. Lett. {\bf 103}, 086403 (2009)}.

\bibitem{carusotto:09}
I. Carusotto, D. Gerace, H. E. Tureci, S. De~Liberato, C. Ciuti, and A.
  Imamo\ifmmode~\check{g}\else \v{g}\fi{}lu,
 Fermionized photons in an array of driven dissipative nonlinear cavities,
 \href{http://dx.doi.org/10.1103/PhysRevLett.103.033601}{%
 Phys. Rev. Lett. {\bf 103}, 033601 (2009)}.

\bibitem{schmid:82}
A. Schmid,
 On a quasiclassical langevin equation,
 \href{http://dx.doi.org/10.1007/BF00681904}{%
 Journal of Low Temperature Physics {\bf 49} (5), 609 (1982)}.

\bibitem{kleinert:95}
H. Kleinert and S. Shabanov,
 Quantum langevin equation from forward-backward path integral,
 \href{http://dx.doi.org/https://doi.org/10.1016/0375-9601(95)00169-4}{%
 Physics Letters A {\bf 200} (3), 224  (1995)}.

\bibitem{Note1}
Note that at finite temperatures, we have $\beta \DOTSB \mapstochar \rightarrow
  \alpha (1 +2 n_B)$ with $n_B = (e^{\hbar \Omega /k_B T} -1)^{-1}$ and we
  obtain the classical limit with $k_B T \gg \hbar \Omega $.

\bibitem{Note2}
Note that the dissipative part of the force can be included via the
  substitution $a\DOTSB \mapstochar \rightarrow a-1$.

\bibitem{Note3}
Since the bifurcation is symmetric in $b$, we simplify the notation by focusing
  on the case $b>0$.

\bibitem{fiedler:08}
B. Fiedler, S. Yanchuk, V. Flunkert, P. H\"ovel, H.-J. W\"unsche, and E.
  Sch\"oll,
 Delay stabilization of rotating waves near fold bifurcation and application to
  all-optical control of a semiconductor laser,
 \href{http://dx.doi.org/10.1103/PhysRevE.77.066207}{%
 Phys. Rev. E {\bf 77}, 066207 (2008)}.

\bibitem{Note4}
We also want to neglect the broadening of the Josephson emission line by
  low-frequency phase noise. It can be neglected, if it is much smaller than
  the linewidth of the resonator $\hbar \Gamma \gg \alpha k_B T $ \cite
  {likharev,Arndt:18}.

\bibitem{likharev}
K. K. Likharev,
 {\em Dynamics of {J}osephson Junctions and Circuits\/}
 (Gordon and Breach Science Publishers, 1986).

\bibitem{Arndt:18}
L. Arndt, A. Roy, and F. Hassler,
 Dual shapiro steps of a phase-slip junction in the presence of a parasitic
  capacitance,
 \href{http://dx.doi.org/10.1103/PhysRevB.98.014525}{%
 Phys. Rev. B {\bf 98}, 014525 (2018)}.

\end{thebibliography}
\end{document}